\documentstyle[epsfig]{aipproc}

\begin{document}
\title{Chaotic Inflation with Variable Space Dimension\thanks{
To appear in the Proceedings of the CAPP2000,
Eds. Ruth Durrer, Juan Garcia-Bellido, and Mikhail Shaposhnikov,
Verbier (Switzerland), July 17-28, 2000.}}

\author{Forough
Nasseri\thanks{e-mail: nasseri@theory.ipm.ac.ir}$^{\dagger}$$^*$ and
Sohrab Rahvar$^{\ddagger}$$^*$}
\address{$^{\dagger}$Physics Department, Shahrood University,
P.O.Box 36155-316, Shahrood, Iran.\\
${^\ddagger}$Physics Department, Sharif University of Technology,
P.O.Box 11365-9161, Tehran, Iran.\\
$^*$Institute for Studies in Theoretical Physics and Mathematics,\\ 
P.O.Box 19395-5531, Tehran, Iran.}

\maketitle

\begin{abstract} 
Assuming the space dimension is not constant but decreases during
the expansion of the Universe, we study chaotic inflation with 
the potential $m^2 \phi^2/2$. We write down field equations in the
slow-roll approximation and define slow-roll parameters by assuming the
space dimension to be a dynamical parameter. The dynamical character of 
the space dimension shifts the initial and final value of the inflaton 
field to larger values, producing delayed chaotic inflation.
We obtain an upper limit for the space dimension at the Planck length.
This result is in agreement with previous works on the effective 
time variation of the Newtonian gravitational constant in a model Universe
with variable space dimensions. We present some cosmological consequences
and calculate observable quantities including the spectral indices, their 
scale-dependence, and the mass of the inflaton field.
\end{abstract}

\section*{Introduction}

In recent work \cite{1} the wave function and the dynamics of a model 
Universe with variable space dimension have been studied in detail. In
this model, the space dimension decreases during the expansion of the 
Universe. There is a dimensional constraint that relates the space
dimension to the size of the Universe. Time variation of the Newtonian
gravitational constant, $G$, is studied in our model Universe \cite{2}.
This study shows an upper limit for the space dimension 
at the Planck length. In \cite{3} we apply the dimensional constraint
of the model to the simplest chaotic inflation and investigate the
dynamical behavior of the inflaton field, the scale factor, the
space dimension, and also their interdependence during the 
inflationary epoch.

Our motivation for studying chaotic inflation with variable space
dimensions is that it enables us to calculate some cosmological 
quantities and compare them with recent observational data, including
the spectral indices. These comparisons indicate to what degree our model 
Universe with variable space dimensions is consistent with the latest
observational data.

Our main results are: {\it i)}
an upper limit for the space dimension at the Planck length,
{\it ii)} a small shift in the value of ${\phi_i}$ and ${\phi_f}$ to
larger values, producing delayed chaotic inflation, and {\it iii)}
derivation of the spectral indices and other observable quantities. As we
show in \cite{3}, in this model the number of space dimensions at the
Planck length must be less than $10$, which is in agreement with the
previous result in \cite{2}.

After this work was substantially complete we learned the picture of
delayed recombination at redshift $z_{\it{rec}} \sim 1000$ caused by early
sources of Ly $\alpha$ radiation \cite{4}. One possibility is that this
delayed recombination may be related to our results of delayed chaotic 
inflation. But it is good science to bear in mind the possibility that
Nature is more complicated than our ideas, a rule that has particular 
force in cosmology because of the limited empirical basis. Future analyses
of {\sc Boomerang} data contains estimates of cosmological parameters and 
informations about the contents and history of the Universe \cite{5}.
If {\sc Boomerang} data does not agree with delayed chaotic inflation,
it would convincingly rule out our model for delayed inflation. Work
on this issue is in progress.

\section*{Universe with Variable Space Dimension}

Mansouri and Nasseri have studied the model Universe with
variable space dimension in detail \cite{1}.
There is a constraint in this model which can be written as 
\begin{equation}
\label{1}
\frac{1}{D}=\frac{1}{C}\ln \left( \frac{a}{a_0} \right)+
\frac {1}{D_0}.
\end{equation}
Here $a$ is the scale factor of Friedmann Universe,
$D$ variable space dimension, and $C$ a constant of the model.
The zero subscript in any quantity e.g. in $a_0$ and $D_0$ denotes
its present value. Choosing the value of the space dimension at the 
Planck epoch $D_P$ we can obtain the value of $C$ by Eq. (\ref{1}).
The dynamical behavior of the Universe in this model was obtained
by using Hawking-Ellis action of a prefect fluid \cite{1}.
 
\section*{Chaotic Inflation with Variable Space Dimension}

Let us now study the inflationary cosmology when the space dimension
is a dynamical parameter. To do this, we first write down our formulation
for general potential, and then use our results in $m^2 \phi^2/2$
potential. The crucial equations are obtained by using chaotic inflation
scenario into our variable space dimension model. Using slow-roll
approximation, we obtain the dynamical equation for the space dimension
\cite{3}
\begin{equation}
\label{2}
{\dot D}^2 \simeq \frac{8 \pi D^3 m^2 \phi^2}{C^2 (D-1) M_P^2}.
\end{equation}
To study the chaotic inflation with variable space dimension, we 
calculate the number of e-foldings ${\cal{N}}$. We find the value of
${\cal{N}}$ by numerical calculations. The initial size of the Universe 
$a_i$ at the beginning of inflation is given by $a_i=a_f\exp(-{\cal{N}})$, 
where $a_f$ is the scale factor at the end of inflation.    
Using dynamical equations, we obtain the value of the inflaton
field at the beginning $\phi_i$ and at the end of inflation $\phi_f$. 
Figure 1 and 2 show $\phi(t)$ as a function of $\ln(a/a_i)$ and $D(t)$,
respectively.  

\begin{figure}[b!] 
\centerline{\epsfig{file=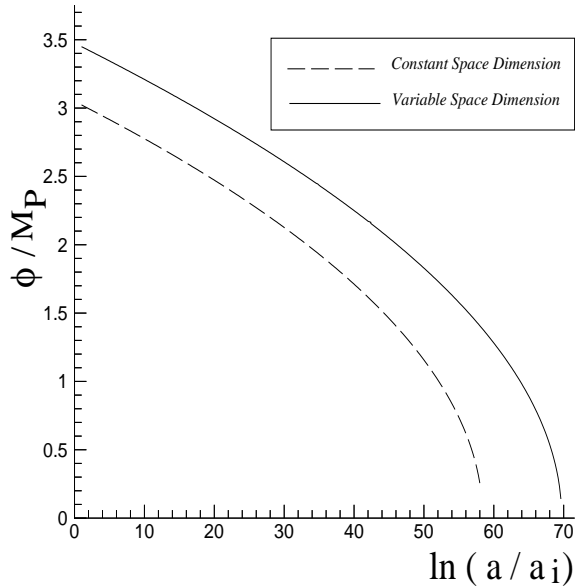,height=3.5in,width=3.5in}}
\vspace{10pt}
\caption{$\phi(t)/M_P$ as a function of $\ln(a/a_i)$ during the
inflationary epoch. The dashed line is for three-space with 
$a_i \simeq 1.87 \times 10^{-25}$ cm and the solid line is for the case
of variable space dimension with $D_P=4$ and $a_i \simeq 2.32 \times
10^{-30}$ cm.} 
\label{fig1}
\end{figure}

\begin{figure}[b!] 
\centerline{\epsfig{file=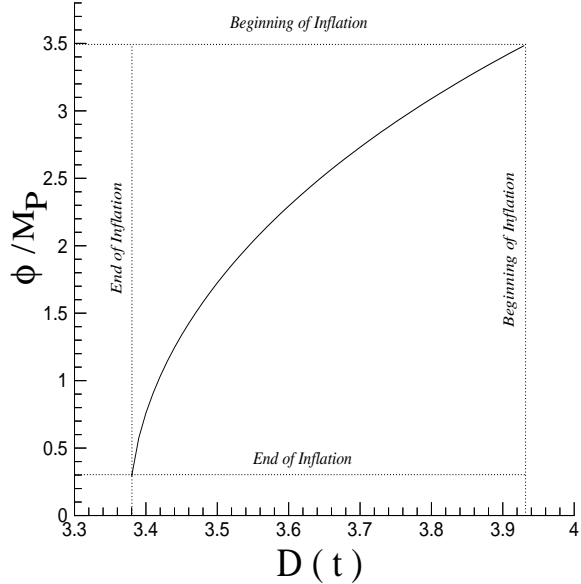,height=3.5in, width=3.5in}}
\vspace{10pt}
\caption{$\phi(t)/M_P$ as a function of $D(t)$ for $D_P=4$ during the 
inflationary epoch.}
\label{fig2}
\end{figure}

\section*{Results}

To obtain the number of e-foldings ${\cal{N}}$, we repeat the analysis of
the inflationary cosmology in three-space. For the space dimension at the
Planck length $D_P=4, 10, 25$ we get ${\cal{N}} \sim 70, 98, 114$,
respectively. The dynamical character of the space dimension increases the
number of e-foldings (for constant three-space we have 58 e-foldings).
Using these e-foldings, we obtain the initial and final value of the
inflaton field. There is a small shift in the value of $\phi_i$ and
$\phi_f$ to larger values. In three-space, inflation ends at
$\sim 10^{-37}$sec. We show the end of inflation is $\sim 10^{-36}$
sec by assuming the space dimension is a dynamical parameter. Therefore,
we conclude that there is a delayed chaotic inflation with variable
space dimension. Using the number of e-foldings, we also obtain the size
of the Universe at the beginning of inflationary epoch. For $D_P=10, 25$
inflation starts from an initial size less than the Planck length. We
know that in three-space the initial size of the inflationary epoch is 
of the order of the Planck length. Particularly, for $D_P=25$ the initial 
size is less than the minimum size of the model, which is $\delta$.
So, we rule out the cases of $D_P \geq 10$ in the model and conclude 
an upper limit for the space dimension at the Planck length, $D_P < 10$.
Our results for an upper limit for the space dimension at the Planck
length is in agreement with previous result in \cite{2}.

\end{document}